\documentclass[11pt,a4paper]{article}
\usepackage[utf8]{inputenc}
\usepackage[T1]{fontenc}
\usepackage{amsmath}
\usepackage{cite}
\usepackage{jheppub} % loads affiliation commands
\usepackage{xcolor}
\definecolor{seagreen}{rgb}{0.05, 0.65, 0.20}

\title{Precision gravity constraints on large dark sectors}

\author[a,b]{Christopher Ewasiuk and Stefano Profumo}

\affiliation[a]{Department of Physics, University of California, Santa Cruz, Santa Cruz, CA, 95064, USA}
\affiliation[b]{Stanford Institute for Theoretical Physics, Department of Physics, Stanford University, Stanford, CA, 94305, USA}

\abstract{
General relativity, treated as a low-energy effective field theory, predicts quantum corrections to Newton’s law of gravitation arising from loops of matter and graviton fields. While these corrections are utterly negligible for the Standard Model particle content, the situation changes dramatically in the presence of a hidden or dark sector containing a very large number of light degrees of freedom. In such cases, loop-induced modifications to the Newtonian potential can accumulate to levels testable in laboratory and astrophysical probes of gravity at short distances. 
In this work we systematically derive and constrain the impact of large dark sectors on precision tests of Newton’s law, translating effective-field-theory predictions into the experimental language of Yukawa-type deviations and inverse-square-law deformations. By mapping precision fifth–force constraints onto bounds on species multiplicities and masses, we show that current and forthcoming experiments already impose nontrivial constraints on the size and structure of hidden sectors coupled only gravitationally. For truly massless hidden states, present data still permit multiplicities as large as $\mathcal{O}(10^{61})$, with modest spin dependence; for finite masses the constraints reduce to the familiar short-range Yukawa parameterization.
Our results provide a model-independent framework for confronting dark-sector scenarios with precision gravity data and clarify how non-minimal scalar couplings, potential higher-derivative poles at large species number, and Kaluza–Klein towers fit within this picture. The approach is complementary to cosmological probes: Big Bang Nucleosynthesis and the Cosmic Microwave Background constrain relic \emph{abundances} under specified production histories, whereas laboratory tests constrain the \emph{spectrum} of light states irrespective of their cosmological population.
}

\begin{document}
\maketitle

\section{Introduction}

Einstein’s general relativity (GR) continues to be the most successful classical description of gravity, having passed all experimental tests from the Solar System to binary pulsars with remarkable accuracy. However, as is by now well recognized, GR should be regarded as a low-energy effective field theory (EFT), valid below the Planck scale $M_{\rm Pl}$, where quantum effects become non-negligible~\cite{Donoghue:1994dn}. Within this EFT framework, quantum loops of matter and gravitons induce calculable corrections to the Newtonian potential, with leading terms scaling as $1/r^3$ at large distances~\cite{Bjerrum-Bohr:2002gqz,Holstein:2023qcy,Faller:2007sy}. The formalism of covariant nonlocal form factors, developed through the Barvinsky--Vilkovisky heat-kernel expansion, provides a powerful language to encode such quantum corrections~\cite{Codello:2015oqa,Codello:2015mba}.

Historically, the search for modifications to Newton’s law at short distances has been motivated by a variety of considerations, from the possibility of new light scalar mediators and axionlike particles, to the presence of large extra dimensions~\cite{Adelberger:2006dh,Kapner:2006si,Hoyle:2004cw}. Precision ``fifth–force’’ experiments, such as torsion-balance measurements pioneered by the Eöt-Wash group, now probe gravitational interactions down to tens of microns~\cite{Tan:2019vwu}. These experiments are typically interpreted in terms of Yukawa-type deviations from the inverse-square law, parametrized by a strength $\alpha$ and range $\lambda$~\cite{Tan:2019vwu,Adelberger:2006dh}. The resulting exclusion curves are among the strongest probes of new macroscopic forces, complementing astrophysical and cosmological constraints.

In the context of hidden or dark sectors, quantum gravitational corrections acquire a qualitatively new role. A single new scalar, fermion, or vector state contributes negligibly to deviations from Newton’s law. However, a sector containing a large multiplicity $N$ of light degrees of freedom can lead to cumulative loop corrections that become appreciable. Massless dark particles (e.g.\ ``dark photons’’) generate enhanced $1/r^3$ power-law corrections, while massive states contribute Yukawa-suppressed tails whose collective strength grows with $N$~\cite{Burns:2014bva,Frob:2016lbc}. This mechanism links the long-standing program of precision gravity tests with the particle-physics motivation to explore large hidden sectors, such as those emerging in dark QCD-like theories or string compactifications.

The central aim of this work is to systematically connect effective field theory predictions for loop-induced gravitational corrections with precision experimental limits, thereby deriving robust bounds on the multiplicity and mass spectrum of hidden sectors coupled only gravitationally. We show how existing torsion-balance data and related measurements can be reinterpreted as constraints on the number of additional massless and light massive fields, yielding quantitative limits that in some cases reach ${\cal O}(10^{61})$ for truly massless species. This approach complements cosmological probes of relativistic degrees of freedom, and highlights an underappreciated synergy between laboratory gravity experiments and hidden-sector model building.

The paper is organized as follows. In Sec.~\ref{sec:quantumcorrections} we review the EFT framework for quantum gravitational corrections and present the relevant nonlocal form factors. Sec.~\ref{sec:massless} discusses the massless limit consistency conditions. In Sec.~\ref{sec:constraints} we apply the formalism to hidden sectors with many degrees of freedom, mapping the loop-induced corrections onto the Yukawa template employed in experiments. In Sec.~\ref{sec:mapping_ff_to_N} we translate published fifth–force bounds into constraints on multiplicities of dark-sector states. We conclude in Sec.~\ref{sec:discussionsconclusions} with a discussion of the implications for dark-sector phenomenology and prospects for future improvements.

\section{Quantum Corrections to Newton’s Constant from Massive and Massless Fields in Effective Field Theory}\label{sec:quantumcorrections}

\subsection{Introduction and Framework}

General relativity can be viewed as a low-energy effective field theory (EFT), valid far below the Planck scale. In this approach, quantum loops of matter fields and gravitons lead to scale-dependent modifications of Newton’s constant. These effects can be described either as a running gravitational coupling $G(q^{2})$ in momentum space, or equivalently through nonlocal form factors multiplying curvature terms in the effective action. At low energies the universal content is carried by nonanalytic terms, such as $q^{2}\ln(1+q^{2}/m^{2})$ for massive particles or $q^{2}\ln(-q^{2})$ for massless fields. The Fourier transforms of these terms produce observable corrections to the Newtonian potential, modifying the familiar $\frac{1}{r}$ force law at short or long distances depending on the particle content.

The subsections that follow outline this framework in detail, first reviewing the structure of the one-loop effective action and then showing how different particle species contribute to the running of G and to corrections of the potential.

\subsection{Effective Action and Nonlocal Form Factors}
\label{sec:eff_action_formfactors}

Quantum corrections to gravity can be systematically treated within the one-loop effective action, viewing general relativity as a low-energy EFT. For a matter field of mass $m$ and spin $s$ one writes \cite{Codello:2015mba,Codello:2015oqa}
\begin{equation}
\Gamma^{(1)}_s \;=\; \tfrac{1}{2}(-1)^{2s}\,\mathrm{Tr}\,\ln\!\big(\Delta_s+m^2\big),
\end{equation}
where $\Delta_s$ is the kinetic operator appropriate to spin $s$. The nonlocal terms of interest appear as form factors multiplying curvature invariants. It is convenient and complete to organize the result in the quadratic curvature basis (see, e.g., \cite{Barvinsky:1990PT1,Donoghue:1994Player,Donoghue:2017EFTreview,Burns:2014FormFactors,Frob:2015Nonlocal}):
\begin{equation}
\Gamma^{(2)}_{\rm matter}
=\int d^4x\,\sqrt{-g}\Big[
R\,F_{1,s}\!\Big(\tfrac{\Box}{m^2}\Big)R
+ R_{\mu\nu}\,F_{2,s}\!\Big(\tfrac{\Box}{m^2}\Big)R^{\mu\nu}
\Big],
\label{eq:quad_action}
\end{equation}
with the dimensionless argument
\begin{equation}
\tau \;\equiv\; -\,\frac{\Box}{m^2}.
\end{equation}
To leading nonlocal order, each spin-$s$ species contributes the universal shape
\begin{equation}
F_{i,s}(\tau)
= c_{i,s}\Big[\,1-\tau\ln\!\Big(1+\frac{1}{\tau}\Big)\Big],
\qquad i=1,2,
\label{eq:Fis_shape}
\end{equation}
which resums the infrared logarithms and smoothly interpolates between the massive and massless regimes.

Note that in the background–field quantization of a (Stueckelberg-regulated) Abelian vector in a covariant gauge, the one-loop effective action at quadratic order reads schematically
\begin{equation}
\Gamma^{(1)}_{(1)} \;=\; \tfrac{1}{2}\,\mathrm{Tr}\ln\!\big(\Delta_{\mu}{}^{\nu}+m^2\big)
\;-\; \mathrm{Tr}\ln\!\big(\Delta_{\rm gh}+m^2\big)
\;+\; \tfrac{1}{2}\,\mathrm{Tr}\ln\!\big(\Delta_{\rm gf}+m^2\big),
\label{eq:ghost_sign_contr}
\end{equation}
where the first term comes from the gauge field, the second from the complex Grassmann Faddeev–Popov ghosts (hence the overall minus sign), and the last from the gauge-fixing Jacobian.\!
Decomposing into spin projectors and rewriting the result in the curvature basis gives
\[
\Gamma^{(2)}_{\rm matter}\supset\!\int\!\sqrt{-g}\,\Big[
R\,F_{1,1}\!\Big(\tfrac{\Box}{m^2}\Big)R
\;+\;
R_{\mu\nu}\,F_{2,1}\!\Big(\tfrac{\Box}{m^2}\Big)R^{\mu\nu}
\Big],
\]
with the same nonlocal shape for both channels,
\(
F_{i,1}(\tau)=c_{i,1}\,[\,1-\tau\ln(1+1/\tau)\,]
\),
but with spin-1 coefficients \(c_{i,1}\) that include the ghost/gauge-fixing contributions.
The static inverse–square–law combination probed by our fits is
\(
F^{\rm(ISL)}_{1}(\tau)=F_{2,1}(\tau)-\tfrac{1}{2}F_{1,1}(\tau)
\),
so the net coefficient is
\(
c_1 \equiv c_{2,1}-\tfrac{1}{2}c_{1,1}.
\)
Including the Faddeev–Popov sector yields a \emph{negative} value for this combination,
\(
c_1=-1/(96\pi^2)
\); see, e.g., \cite{Barvinsky:1990PT1,Burns:2014FormFactors,Frob:2015Nonlocal}
for explicit derivations in the covariant form-factor language.

It is important to emphasize that this curvature--squared form--factor
coefficient $c_{1}$ is not identical to the coefficient of the spin--1
contribution in the long--distance potential written in Eq.~(I.1) of
Ref.~\cite{Burns:2014FormFactors}, where only the \emph{physical} vector
degrees of freedom contribute (counted by $N_{1}$), with the ghost sector
omitted from that parametrization.  After translating between the physical
field--counting basis used in Eq.~(I.1) and the covariant form--factor basis
used here (which automatically incorporates the ghost sector via
Eq.~\eqref{eq:ghost_sign_contr}), both approaches yield the same long--distance
$1/r^{3}$ correction to the Newtonian potential.  The negative value
\eqref{eq:cs_values} therefore reflects the standard background--field definition
of the ISL coefficient for spin--1 matter.%

\medskip\noindent
%\textit{Observable combination for inverse–square–law tests.}
At the linearized level about flat space, the static potential probes a definite combination of the spin-2 and spin-0 channels carried by the propagator. This can be expressed as
\begin{equation}
F^{\rm(ISL)}_{s}(\tau)
= F_{2,s}(\tau)-\tfrac{1}{2}F_{1,s}(\tau)
= c_s\Big[\,1-\tau\ln\!\Big(1+\frac{1}{\tau}\Big)\Big],
\qquad
c_s \equiv c_{2,s}-\tfrac{1}{2}c_{1,s}.
\label{eq:Fs_ISL}
\end{equation}
Equation \eqref{eq:Fs_ISL} shows explicitly that both $R^2$ and $R_{\mu\nu}R^{\mu\nu}$ are generated at one loop, while clarifying why a single coefficient $c_s$ suffices for the static potential: it is the short-distance combination entering inverse–square–law (ISL) observables.

\medskip\noindent
%\textit{Explicit coefficients and multiplicities.}
For minimally coupled matter fields, the coefficients entering the ISL combination are
\begin{equation}
c_0 = \frac{1}{384\,\pi^2}\ \ \text{(real scalar)},\quad
c_{1/2} = \frac{1}{192\,\pi^2}\ \ \text{(Dirac fermion)},\quad
c_1 = -\,\frac{1}{96\,\pi^2}\ \ \text{(vector+ghosts)}.
\label{eq:cs_values}
\end{equation}
Summing over all species gives
\begin{equation}
F^{\rm(ISL)}(\tau)\;=\;\sum_{s} N_s\,c_s\,
\Big[\,1-\tau\ln\!\Big(1+\frac{1}{\tau}\Big)\Big],
\label{eq:sum_species}
\end{equation}
where $N_s$ counts the number of fields of spin $s$. If a non-minimal scalar coupling $\tfrac{1}{2}\xi R\phi^2$ is present, the trace (spin-0) channel carries the familiar $(1-6\xi)^2$ weight \cite{Barvinsky:1990PT1,Burns:2014FormFactors}; in practice one may track this by $c_0\!\to c_0(1-6\xi)^2$ in \eqref{eq:sum_species}.

\medskip\noindent
%\textit{Asymptotics and connection to the potential.}

The long--distance $1/r^{3}$ correction to the Newtonian potential originates
from the nonanalytic part of the covariant form factors $F_{i,s}(\tau)$.  In
the massless or high--momentum regime ($\tau\to\infty$), the form factors take
the well--known limit $F_{i,s}(\tau)\simeq c_{i,s}(\ln\tau-2)$, reproducing the
$\ln(-\Box/\mu^{2})$ running characteristic of massless loops
\cite{Barvinsky:1990PT1,Donoghue:1994Player}.  For massive fields with
momenta well below threshold ($|\tau|\ll1$), the same nonlocal structure
admits the local curvature expansion
\begin{equation}
F_{i,s}(\tau)=c_{i,s}\Big(\tfrac{\tau}{6}-\tfrac{\tau^{2}}{60}+\cdots\Big),
\end{equation}
which is the low–momentum expansion of the form factor that yields the
universal $1/r^{3}$ tail upon resummation.  At intermediate distances,
this physics is conveniently interpolated by a Yukawa--envelope fit
\cite{Frob:2015Nonlocal,Burns:2014FormFactors}.  Our mapping from ISL data to
constraints on hidden fields therefore consistently incorporates both the
curvature--squared operators arising at low momenta and their nonlocal
running inherited from the full form factor.

% Replaced the text below
% In the massless limit ($\tau\to\infty$), $F_{i,s}(\tau)\simeq c_{i,s}(\ln\tau-2)$, reproducing the standard running $\ln(-\Box/\mu^2)$ \cite{Barvinsky:1990PT1,Donoghue:1994Player}. For $|\tau|\ll1$ (momenta well below $m$),
% \begin{equation}
% F_{i,s}(\tau)=c_{i,s}\Big(\frac{\tau}{6}-\frac{\tau^2}{60}+\cdots\Big),
% \end{equation}
% which yields the local derivative series generating the universal long-distance $1/r^3$ tail in the static potential. At intermediate distances the same physics is conveniently encoded by a Yukawa-envelope fit \cite{Frob:2015Nonlocal,Burns:2014FormFactors}. In this way, our mapping from ISL data to bounds on the cumulative contribution of hidden fields naturally accounts for both curvature–squared operators and their nonlocal running.
%
%

%
%================================================%===
%
%
%The nonlocal term above leads to $1/r^3$-type corrections in the gravitational potential. 
%
For example, integrating out a real scalar yields a correction
\begin{equation}
U(r) \;=\; -\frac{G m_1 m_2}{r}\left[\,1 + \frac{c_0}{G} \frac{1}{r^2} + \ldots \,\right],
\end{equation}
where the second term encodes the leading quantum effect. In general summing over all particle species (weighted by $c_s$) determines the net sign and magnitude of these corrections.

If the loop particle is massless, thus $\tau \to \infty$,  the form factors reduce to the pure nonanalytic structure $F_s(\tau) \sim c_s \ln(-\Box/\mu^2)$, with $\mu$ a renormalization scale. This directly implies a scale-dependent gravitational coupling $G(q^2)$. We discuss the massless case in more detail in the following section. For particles with nonzero mass, the mass scale $m$ acts as a threshold: at distances much larger than $1/m$, the quantum corrections are strongly suppressed, while at shorter distances the corrections gradually approach the same logarithmic running seen in the massless case. In this way, the massive case smoothly connects between $1/m^2$-suppressed effects in the low energy ($q^2 \ll m^2$) and universal logarithmic behavior in the high energy case ($q^2 \gg m^2$), a feature that plays an important role in precision tests of gravity across laboratory and astrophysical regimes.

\subsection{Running and Threshold Effects in $G(q^2)$}

The one-loop corrections can be interpreted as being momentum-dependent, or ``running,'' Newton’s constant. In flat-space, the running of Newton’s constant can be parametrized by a correction to its
inverse propagator,
\begin{equation}
G_{\text{eff}}^{-1}(q^2) \;=\; G_0^{-1}\left[1 + \delta_s(q^2)\right],
\end{equation}
where $G_0$ is the noncorrected gravitational constant and $\delta_s(q^2)$ encodes the quantum correction from a particle of mass $m$ and spin $s$. In the case of massless fields, $m \to 0$, the expression reduces to a purely nonlocal form,
\begin{equation}
\delta_s(q^2) \;\sim\; c'_s\, G_0 q^2 \ln\!\left(\frac{-q^2}{\mu^2}\right),
\end{equation}
where $\mu$ is a renormalization scale. This result is explicitly derived and discussed in detail later in Section \ref{sec:massless}. This universal logarithmic dependence is responsible for long-distance, nonlocal corrections to Newton’s law~\cite{Donoghue:1994dn,Bjerrum-Bohr:2002gqz,Khriplovich:2002bt}.

For massive fields, the behavior is governed by threshold effects. At low momentum, $q^2 \ll m^2$, the logarithm can be expanded and the correction takes the form
\begin{equation}
\delta_s(q^2) \;\approx\; \frac{c_s}{2} \, G_0 \, \frac{q^4}{m^2},
\end{equation}
which is analytic in $q^2$ and therefore indistinguishable from local higher-derivative operators. For graviton propagators with large momentum transfer, $q^2 \gg m^2$, the particle contributes fully to the running and the correction asymptotes to
\begin{equation}
\delta_s(q^2) \;\approx\; c_s\, G_0\, q^2 \ln\!\left(\frac{q^2}{m^2}\right).
\end{equation}
The full nonlocal expression, $\ln(1 + q^2/m^2)$, thus interpolates smoothly between these two regimes, ensuring that heavy particles decouple at low energies while recovering the massless logarithmic running in the ultraviolet.

Summing over all particle species yields a compact expression for the effective coupling in Euclidean space,  
\begin{equation}
G_\mathrm{eff}(-q^2) \;\simeq\; G_0\left[\,1 + G_0 q^2 \sum_s c_s \ln\!\left(1+\frac{q^2}{m_s^2}\right) + \ldots \,\right],
\end{equation}
where the sum runs over all fields with mass $m_s$.  
This interpolating behavior between the low-energy analytic regime and high-energy logarithmic running captures the essential physics of how massive fields contribute to gravitational quantum corrections. The threshold scale $m$ determines whether particles contribute significantly to the running of Newton's constant: light particles ($q^2 \gg m^2$) participate fully in the quantum corrections, while heavy particles ($q^2 \ll m^2$) decouple and contribute only suppressed analytic terms. This mass-dependent decoupling will prove crucial when analyzing the cumulative effects of large hidden sectors, as it determines which mass ranges are most relevant for observable deviations from Newton's law at different distance scales.

\subsection{Correction to the Newtonian Potential}
As outlined above, the corrections to the Newtonian potential are obtained by integrating the nonlocal form factor above for all possible particle spins $s$. In the case of the massless fields, such corrections to the gravitational potential induce long range effects of the form $O(1/r^3)$~\cite{Donoghue:1994dn,Bjerrum-Bohr:2002gqz,Khriplovich:2002bt,Faller:2007sy,Holstein:2023qcy}:

\begin{equation}
U(r) = -\frac{G m_1 m_2}{r} \left[ 1 + \alpha\, \frac{G}{r^2} + \ldots \right]
\end{equation}
where ~\cite{Bjerrum-Bohr:2002gqz}
\begin{equation}
\alpha = \frac{41}{10\pi}\qquad \text{(graviton loops)},
\end{equation}
is obtained via graviton ($s = 2$) corrections and is modified via additional contributions from spins 0, $1/2$, 1 to get:
\begin{equation}
\alpha = \frac{41}{10\pi} + \sum_f N_f \frac{1}{20\pi} + \sum_s N_s \frac{1}{120\pi} + \sum_v N_v \left( -\frac{1}{10\pi} \right)
\end{equation}
where $N_{f,s,v}$ count Dirac fermions, real scalars, and vectors. For massive fields, corrections are exponentially suppressed beyond the Compton wavelength and interpolate between short-distance (massless) and long-distance (suppressed) regimes. Explicit analytic and numerical forms are given in~\cite{Frob:2016lbc,Burns:2014bva}, including Bessel and Struve functions. 

%============================================
\subsection{Validity of the EFT and species effects.}
Loop-generated quadratic curvature terms can be reorganized into massive spin-0 and spin-2 propagating poles (the latter being a ghost in a purely local $R^2\!+\!R_{\mu\nu}R^{\mu\nu}$ truncation). In a Wilsonian picture the induced masses scale schematically like
\[
M_{0,2}\;\sim\;\frac{M_{\rm Pl}}{\sqrt{\alpha_{0,2}}}\,,
\qquad
\alpha_{0,2}\;\propto\;\frac{N}{16\pi^2},
\]
so extremely large $N$ can lower $M_{0,2}$. Our laboratory analysis is performed at distances $r_\star$ in the mm–$\mu$m range. If $M_{0,2}^{-1}\ll r_\star$, the massive poles merely renormalize the local curvature terms and the familiar nonlocal $1/r^3$ tail governs the signal. If instead $M_{0,2}^{-1}\gtrsim r_\star$, the same physics manifests as \emph{additional Yukawa components} with ranges $\lambda_{0,2}\sim M_{0,2}^{-1}$, which are already encompassed by our envelope fit and lead to \emph{tighter} constraints. Thus a conservative consistency condition for our nonlocal description is $M_{0,2}^{-1}\ll r_\star$; when it fails, one should include the corresponding Yukawa piece explicitly—which our mapping accommodates without changing the empirical logic of the bounds.

%============================================

\section{Massless Limit Consistency}\label{sec:massless}

A crucial consistency check for the quantum corrections derived in the preceding sections is that, in the limit of vanishing mass, the expressions for massive fields smoothly reproduce the well-known massless results both in momentum-space running and in the real-space potential. This property is explicitly realized in the referenced calculations~\cite{Frob:2016lbc, Burns:2014bva}.

\subsection{Momentum-Space Running}

For a field of mass $m$ and spin $s$, the nonlocal correction to Newton’s constant in momentum space is obtained via integrating the quantum corrections given by $F_s$ in Eq. \ref{eq:quantum_corr} to obtain the form ~\cite{Frob:2016lbc}:
\begin{equation}
    \delta_s(q^2) = c_s\, G_0\, q^2 \ln\left( 1 + \frac{q^2}{m^2} \right).
\end{equation}
In the massless limit, we are interested in the $m \to 0$ limit. Doing so, the logarithm expands as:
\begin{equation}
    \ln\left(1 + \frac{q^2}{m^2}\right) \longrightarrow \ln\left( \frac{q^2}{m^2}\right).
\end{equation}
With this minor expansion, the overall expression for the massless correction to Newtons constant takes the form
\begin{equation}
    \delta_s(q^2) \to c_s\, G_0\, q^2 \ln\left( \frac{q^2}{\mu^2} \right).
\end{equation}
 where $\mu$ has replaced the mass $m$ and defines the renormalization scale in the massless case. This matches the standard massless result, where the running of $G$ is governed by a $q^2 \ln(-q^2)$ form factor~\cite{Donoghue:1994dn, Bjerrum-Bohr:2002gqz}. In the limit $m \rightarrow 0$, the nonlocal form factor reduces to the universal massless expression $ \sim q^2 ln(-q^2)$, in agreement with earlier EFT analyses of graviton and massless matter loops. The scale m in the massive case thus plays the role of a regulator, which is replaced by a renormalization scale $\mu$ in the massless theory.

\subsection{Position-Space Potential}

\begin{center}
\begin{tabular}{|l|l|l|}
    \hline
    \textbf{Massive Correction} & \textbf{$m \to 0$ Limit (Massless)} & \textbf{Source} \\
    \hline
    Exponential/Yukawa: $e^{-mr}/r^3$ & Nonanalytic $1/r^3$ tail & \cite{Frob:2016lbc,Burns:2014bva} \\
    Bessel/Struve function form & Expands to $1/r^3$ & \cite{Burns:2014bva} \\
    \hline
\end{tabular}
\end{center}
The effect of quantum corrections can also be understood directly in position space by examining their contribution to the Newtonian potential. For a particle of finite mass $m$, the correction takes a Yukawa-suppressed form
\begin{equation}
    \delta U(r) \;\sim\; \frac{e^{-mr}}{r^3},
\end{equation}
so that the contribution is short-ranged and becomes negligible beyond the particle’s Compton wavelength. More detailed treatments~\cite{Burns:2014bva} show that the exact expression can be written in terms of special functions (e.g., Bessel and Struve), but the essential behavior is well captured by the exponential factor.  Both~\cite{Burns:2014bva,Frob:2016lbc} express the full quantum-corrected potential $V(r)$ for massive loops using Bessel/Struve functions:
\begin{itemize}
    \item For $mr \ll 1$, the expansion reproduces the $1/r^3$ behavior of massless loop corrections.
    \item For $mr \gg 1$, the correction is exponentially (Yukawa) suppressed.
\end{itemize}

In the massless limit, $m \to 0$, the exponential suppression goes to unity, leaving a long-range correction of the form
\begin{equation}
    \delta U(r) \sim \frac{1}{r^3}.
\end{equation}
This result matches the standard nonanalytic correction derived in effective field theory and demonstrates the smooth connection between the massive and massless cases. At short distances ($mr \ll 1$) the potential reduces to the $1/r^3$ form characteristic of massless loops, while at large distances ($mr \gg 1$) the correction is exponentially suppressed and therefore negligible. In this way, the position-space picture confirms the consistency of the massless limit and illustrates how heavy particles decouple from long-range gravitational physics.

\section{Constraints on Hidden or Dark Sectors with Many Degrees of Freedom}\label{sec:constraints}

The preceding analysis, based on the Standard Model particle content, shows that quantum corrections to Newton's law are utterly negligible at accessible length scales. However, the situation can change qualitatively in the presence of a dark sector containing a large number of light or very weakly coupled particles---for example, many real scalars, fermions, or vectors with masses $m_i$ below or near the meV scale.

\subsection{Quantum Corrections from Large Hidden Sectors.}
The full quantum-corrected Newtonian potential arises from the combined effect of all light and heavy degrees of freedom running in loops ~\cite{Bjerrum-Bohr:2002gqz,Frob:2016lbc}. Each particle species contributes a correction determined by its spin, mass, and multiplicity, with massless states generating long-range, nonanalytic terms and massive states producing short-range Yukawa-suppressed effects. In practice, the net modification is obtained by summing over all particle types, weighted by the spin-dependent coefficients introduced in the previous section. This leads to the general expression for a modified gravitational potential:

\begin{equation}
U(r) = -\frac{G m_1 m_2}{r} \Bigg[
1
+ \sum_{j\in \text{massless}} \alpha_j \frac{G}{r^2}
+ \sum_{i\in \text{massive}} \beta_i \frac{G e^{-2 m_i r}}{r^2}
+ \cdots
\Bigg]
\end{equation}
with $\alpha_j$ and $\beta_i$ determined by the spin-dependent loop coefficients $c_s$ (see previous section), and the sums extend over all massless and massive dark sector states.

In the massive case with $N$ nearly degenerate, weakly-coupled species with similar masses $m_X$, the Yukawa correction is enhanced by the total number of additional degrees of freedom $N$,
\begin{equation}
U_\text{Yukawa}(r) \sim -\frac{G^2 m_1 m_2}{r} \, N\, \beta_X \frac{e^{-2 m_X r}}{r^2}.
\label{eq: Yukawa}
\end{equation}
This means that the quantum gravitational correction, while tiny for each individual state, can become collectively significant if $N$ is large or if $m_X$ is sufficiently small ($m_X \lesssim$ meV). For heavier states the Yukawa corrections are strongly suppressed, but ultralight particles can give rise to measurable deviations in precision experiments.

%\paragraph{Which states contribute at low energy?}
Note that our reinterpretation counts any \emph{propagating} degree of freedom lighter than the experimental lever arm. Concretely, a state contributes to the potential if its mass satisfies $m\lesssim r_\star^{-1}$. In confining theories, heavy resonances above this scale are integrated out and do not affect our fits, whereas light PNGBs, glueballs, or other composites couple universally to gravity and contribute on the same footing as elementary fields. Our constraints apply to any light degree of freedom that behaves as an
effectively pointlike scalar at the momentum transfers relevant for ISL
experiments.  For PNGBs, this condition is naturally satisfied because their
masses can be parametrically smaller than their compositeness scale.  For
glueballs, this occurs only if the confinement scale of the hidden sector is
sufficiently low (e.g.\ keV–MeV), so that the lightest glueball is both
kinematically accessible and structureless at the probed momenta.  In
contrast, QCD glueballs are too heavy to satisfy this condition, and do not
enter our bounds. The bounds therefore depend only on the infrared spectrum, not on UV compositeness.

\subsection{Experimental Limits: Mapping onto the Yukawa Template.}
A standard approach to constraining possible deviations from Newton’s law is to express the potential in terms of a Yukawa-like correction, parameterized as

\begin{equation}
U_{\text{exp}}(r) = -\frac{G m_1 m_2}{r} \left[ 1 + \gamma\, e^{-r/\lambda} \right]
\end{equation}
where $\gamma$ is the dimensionless strength and $\lambda$ the range of a hypothetical new Yukawa force. This form provides a convenient benchmark because it captures both the possibility of a new long-range force (if $\lambda$ is large) and a short-range modification (if $\lambda$ is small). Comparing to the Yukawa correction in Eq. \ref{eq: Yukawa}, the collective quantum loop correction from a large hidden sector maps onto this template by identifying
\begin{equation}
\gamma = N\, \beta_X \frac{G}{2 m_X \lambda} \qquad\text{with}\qquad \lambda = \frac{1}{2 m_X},
\end{equation}
where the detailed normalization depends on the spin, couplings, and explicit computation of $\beta_X$ for the field(s) in question~\cite{Frob:2016lbc, Burns:2014bva}. However, the mapping is not reliable as $m \to 0$; we return to a detailed discussion of this massless limit in Sec.~\ref{sec:mapping_ff_to_N}.

\subsection{Constraints and Projected Sensitivity}

Experimental searches for deviations from Newton’s law are commonly presented as exclusion plots in the $(\gamma,\lambda)$ plane, where $\gamma$ sets the relative strength and $\lambda$ the range of a Yukawa-type correction. Within our framework, these bounds can be directly reinterpreted as limits on the number and properties of hidden-sector states that contribute to quantum loop corrections of gravity.

For instance, identifying $\lambda = 1/(2 m_X)$ for a species of mass $m_X$, one finds that experiments typically constrain $|\gamma| \lesssim 10^{-2}$--$10^{-4}$ at micron length scales~\cite{Burns:2014bva}. If the collective enhancement from a large number of species $N$ or from very light masses $m_X$ pushes the predicted quantum correction above these bounds, the corresponding models can be ruled out or tightly constrained. The overall size and sign of the effect depend on the weighted sum of coefficients $c_s$ for the contributing fields, and can therefore be evaluated model by model using the master expressions for $U(r)$ derived above~\cite{Frob:2016lbc, Bjerrum-Bohr:2002gqz}.
If the hidden sector also contains massless states (such as scalars or dark photons), the cumulative correction is instead of the form
\begin{equation}
    \alpha_{\text{tot}} \;=\; \sum_{j \in \text{massless}} N_j\, c_j,
\end{equation}
leading to an additive $1/r^3$ term in the potential. While these corrections fall off faster than Yukawa-type terms and are generally harder to detect, their cumulative effect could still be constrained through precision fits to short-distance deviations from the $1/r$ force law.
Still, the quantum loop corrections to the gravitational potential---Yukawa-suppressed for each individual massive state and $1/r^3$-suppressed for massless ones---can, if multiplied by a large number of hidden-sector degrees of freedom, become significant enough to be probed by laboratory experiments. The basic procedure is to map the theoretical prediction
\begin{equation}
    \Delta U(r) \;=\; -\frac{G m_1 m_2}{r} 
    \sum_X N_X \, \beta_X \, \frac{e^{-2 m_X r}}{r^2}
\end{equation}
onto the experimental Yukawa form with parameters $(\gamma,\lambda)$. Published exclusion plots in the $(\gamma,\lambda)$ plane then translate directly into limits on the number $N_X$ and mass $m_X$ of new species~\cite{Frob:2016lbc, Burns:2014bva}, as we will see below.

\section{From fifth–force bounds to multiplicity limits}
\label{sec:mapping_ff_to_N}

We now show how experimental fifth–force searches can be recast as direct bounds on the multiplicity of hidden-sector degrees of freedom that couple to gravity. This provides a framework between laboratory constraints on Yukawa or power–law deviations from Newton’s law and theoretical scenarios with large numbers of light or massless states. Throughout this section we set $\hbar = c = 1$ and denote Newton’s constant by $G_N$.

\subsection{Set–up}

To connect experimental fifth–force searches to hidden–sector physics, we consider a generic dark sector containing $N_s$ real scalars, $N_f$ Dirac fermions, and $N_v$ vectors, all coupled only through gravity. The effect of these fields is to modify the Newtonian potential by loop exchange, producing a correction of the form
\begin{equation}
\frac{\Delta U}{U_{\rm Newt}}(r)
=\sum_{X=s,f,v} N_X\,\frac{G_N}{r^2}\,\mathcal{K}_X(m_X r),
\qquad U_{\rm Newt}(r)=-\frac{G_N m_1 m_2}{r},
\label{eq:master}
\end{equation}
where $\mathcal{K}_X$ is a dimensionless loop kernel encoding the spin and mass dependence of the exchanged particle. 
Two limits of $\mathcal{K}_X$ are especially important:
\begin{equation}
\mathcal{K}_X(0)=\beta_X,
\quad
\beta_s=\frac{1}{120\pi},\quad
\beta_f=\frac{1}{20\pi},\quad
\beta_v=-\frac{1}{10\pi},
\label{eq:beta}
\end{equation}
for massless fields, and 
\begin{equation}
\mathcal{K}_X(\xi)\xrightarrow[\xi\gg1]{}\beta_X^{\rm(as)}(\xi)\,e^{-2\xi},\qquad \xi\equiv m_X r,
\label{eq:kernel-as}
\end{equation}
The superscript (as) indicates the asymptotic form: $\beta^{({\rm as})}_X(\xi)$ gives the coefficient in the large-$\xi$ expansion, which generally differs from the massless coefficient $\beta_X$ due to additional corrections that emerge in the massive case at large distances. For massive fields at distances large compared to their Compton wavelength. Thus, massless species always generate a long–range $1/r^3$ correction, whereas massive species decouple exponentially at large $r$.

On the experimental side, deviations from Newton’s law are typically reported in two standard parameterizations:
\begin{equation}
V_k(r)=-\frac{G_N m_1 m_2}{r}\left[1+\beta_k\left(\frac{r_0}{r}\right)^{k-1}\right],
\label{eq:powerlaw}
\end{equation}
for power–law deformations of the inverse–square law, and 
\begin{equation}
V_Y(r)=-\frac{G_N m_1 m_2}{r}\Big[1+\alpha_{\rm exp}(\lambda)\,e^{-r/\lambda}\Big],
\label{eq:yukawa}
\end{equation}
for Yukawa–type forces. The loop–induced $1/r^3$ correction maps onto the $k=3$ case of Eq.~\eqref{eq:powerlaw}, while finite–mass fields map onto the Yukawa form with range $\lambda=1/(2m_X)$. 

Our methodology for constraining multiplicity limits  involves utilizing given experimental exclusion curves to obtain values of $\alpha$ and $\lambda$ to set direct bounds on the number of degrees of freedom. In the following subsections we treat separately the \emph{massless} case, which allows closed–form analytic bounds, and the \emph{massive} case, which requires mapping onto Yukawa envelopes.

\subsection{Massless dark degrees of freedom: closed–form bound with numbers}
When $m_X=0$, Eqs.~\eqref{eq:master} and \eqref{eq:beta} give
\begin{equation}
\frac{\Delta U}{U_{\rm Newt}}(r)
=\Big(\frac{N_f}{20\pi}+\frac{N_s}{120\pi}-\frac{N_v}{10\pi}\Big)\frac{G_N}{r^2}.
\label{eq:frac-massless}
\end{equation}
Direct ISL tests at short range (torsion balances) often report the $k=3$ power–law coefficient at $r_0=1~{\rm mm}$.  Writing the fractional residual as $\varepsilon(r)\equiv |\Delta U/U_{\rm Newt}|$, the $k=3$ parameterization implies
\begin{equation}
\varepsilon(r)\,r^2=\beta_3\,r_0^2\qquad(k=3,\ r_0=1~{\rm mm}),
\label{eq:eps-r2}
\end{equation}
independent of $r$.  Combining Eqs.~\eqref{eq:frac-massless} and \eqref{eq:eps-r2} yields the general multiplicity bound
\begin{equation}
\boxed{\;
\Big|\frac{N_f}{20\pi}+\frac{N_s}{120\pi}-\frac{N_v}{10\pi}\Big|
\;\le\;\frac{\beta_3\,r_0^2}{G_N}\; .\;}
\label{eq:massless-master}
\end{equation}

Using the most precise published ISL \emph{power–law} constraint (Eöt–Wash–style torsion balance),\footnote{Tan \emph{et al.} quote power–law bounds including $k=3$; for convenience we take their tabulated $k=3$ value at 68\%\,CL, $|\beta_3|\lesssim 7.5\times10^{-5}$ \cite{Tan:2019vwu}. Older analyses (e.g.\ \cite{Hoyle:2004cw}) give consistent though weaker values.  At 95\%\,CL, numbers inflate by a factor $\sim\!1.6$--2.}
\begin{equation}
|\beta_3|\ \lesssim\ 7.5\times10^{-5}\quad(68\%~{\rm CL}) \qquad\Rightarrow\qquad
\frac{\beta_3 r_0^2}{G_N}\ \lesssim\ 2.87\times 10^{59},
\end{equation}
where we used $r_0=1~{\rm mm}$ and $G_N=l_P^2=2.611\times10^{-70}\,{\rm m}^2$.
Equation~\eqref{eq:massless-master} then gives the \emph{per–spin} limits (assuming only one spin species is present):
\begin{align}
\boxed{~N_s \ \lesssim\ 1.08\times 10^{62}~} &\qquad\text{(real scalar)}, \label{eq:Ns-num}\\
\boxed{~N_f \ \lesssim\ 1.80\times 10^{61}~} &\qquad\text{(Dirac fermion)}, \label{eq:Nf-num}\\
\boxed{~N_v \ \lesssim\ 9.02\times 10^{60}~} &\qquad\text{(vector; using $|\beta_v|$)}. \label{eq:Nv-num}
\end{align}
For Weyl/Majorana fermions, divide \eqref{eq:Nf-num} by two.  If multiple spins are present, apply \eqref{eq:massless-master} to the spin–weighted sum; one may also impose a robust (no–cancellation) limit by bounding the sum of absolute contributions. Equations \eqref{eq:Ns-num}–\eqref{eq:Nv-num} show that current short–range ISL data allow extremely large multiplicities of truly massless gravitationally–coupled fields; any improvement in $|\beta_3|$ translates linearly to these bounds.

\subsection{Massive dark degrees of freedom: mapping to Yukawa envelopes}
For $m_X>0$, the experimental summaries provide 95\%\,CL limits $\alpha_{\rm exp}^{\max}(\lambda)$ in Eq.~\eqref{eq:yukawa}. Identify the range with particle mass as
\begin{equation}
\lambda=\frac{1}{2m_X},
\label{eq:lambda-mass}
\end{equation}
and match the leading long–range piece of \eqref{eq:master} to the Yukawa form at the apparatus lever arm $r_\star$ (choosing $r_\star\simeq \lambda$ maximizes sensitivity). Defining a dimensionless shape factor $\mathcal{S}_X(\xi)\equiv \mathcal{K}_X(\xi)\,e^{2\xi}$ with $\mathcal{S}_X(0)=1$ and $\mathcal{S}_X(\tfrac12)=\mathcal{O}(1)$, one obtains
\begin{equation}
\alpha_{\rm th}(\lambda)
\simeq N_X\,\beta_X\,\frac{G_N}{r_\star^2}\,\mathcal{S}_X\!\left(m_X r_\star\right)
\quad\Rightarrow\quad
\boxed{\;
N_X^{\max}(\lambda)
\simeq 
\frac{\alpha_{\rm exp}^{\max}(\lambda)}{\beta_X\,\mathcal{S}_X(1/2)}\,
\frac{\lambda^2}{G_N}\;,
\ \ \lambda=\frac{1}{2m_X}.
\;}
\label{eq:Nmax-massive}
\end{equation}
Setting $\mathcal{S}_X(1/2)=1$ gives a slightly conservative estimate that is convenient for quick re-interpretations of published $(\alpha,\lambda)$ curves (torsion balances, LLR, planetary dynamics, \emph{etc.}). Applying the results from precision experiments at varying length scales, we can obtain continuous limits on the multiplicities of dark sectors, as is illustrated in fig.~\ref{fig:N_massive} below. 

\
\paragraph{Including additional Yukawa poles when they enter the lever arm.}
If the higher-derivative poles discussed in Sec.~\ref{sec:quantumcorrections} descend into the laboratory window, their effect is incorporated by \emph{adding} explicit Yukawa pieces to the static potential, rather than relying solely on the single-envelope mapping. We write
\begin{equation}
\label{eq:multiYukawa}
U(r)\;=\;-\frac{G\,m_1 m_2}{r}\Bigg[\,1+\sum_{k=1}^{n_\mathrm{Y}}\alpha_k\,e^{-r/\lambda_k}\Bigg],
\end{equation}
with $(\lambda_k,\alpha_k)$ the range and strength of the $k$-th Yukawa term. For the quadratic-curvature spin channels one may take $n_\mathrm{Y}=2$, with
\begin{equation}
\label{eq:Yukawa0_2_mapping}
(\lambda_0,\alpha_0)\ \ \text{and}\ \ (\lambda_2,\alpha_2),\qquad
\lambda_{0,2}=\frac{1}{2M_{0,2}},
\end{equation}
mirroring the convention in Eq.~\eqref{eq:lambda-mass}. The data analysis then proceeds by profiling the likelihood (or $\chi^2$) over $(\alpha_0,\alpha_2)$ at fixed $(\lambda_0,\lambda_2)$, yielding one–sided bounds $\alpha_{0,2}^{95}(\lambda_{0,2})$ (single–pole case) or joint contours for two poles. Because Eq.~\eqref{eq:multiYukawa} fits the \emph{sum} of deviations from $1/r$, any additional pole within the probed range can only increase the total deviation unless it is finely tuned to cancel other pieces; our envelope construction avoids such tuned cancellations and thus remains conservative.

\begin{figure}[t!]
    \centering
    \makebox[\textwidth][c]{%
        \includegraphics[width=1\linewidth]{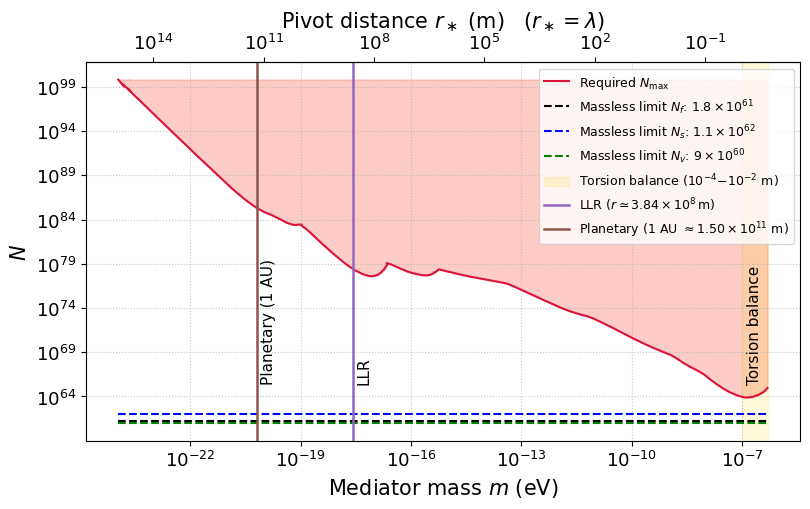}
    }
    \caption{Degrees of freedom set by experimental constraints on $G$ and the Yukawa parameters $(\alpha,\lambda)$. The shaded region indicates the exclusion set by experimental limits at each distance scale $r_\star$.}
    \label{fig:N_massive}
\end{figure}

\subsection{Experimental Multiplicity limits in both the Massive and Massless Case}

Utilizing values of $\alpha$ and $\lambda$ from experimental exclusion curves, we construct analogous exclusion plots for the upper limits on the number of degrees of freedom based on Eq.~\eqref{eq:Nmax-massive} for massive particles in Figure~\ref{fig:N_massive}. For concreteness, we assume the shape factor $S_X$ is unity and model the gravitational correction as arising entirely from a large sector of fermionic particles with spin-dependent loop coefficient $\beta_f = \frac{1}{20\pi}$. The methodology readily extends to scalar and vector particles by substituting the appropriate coefficients $\beta_s = \frac{1}{120\pi}$ and $\beta_v = -\frac{1}{10\pi}$, respectively, which modify the bounds by at most an order of magnitude.

Figure~\ref{fig:N_massive} presents these constraints in the $(m, N)$ plane, where the horizontal axis shows the dark sector particle mass $m$ and the vertical axis displays the maximum allowed multiplicity $N$. The corresponding pivot distance $r_\star = \lambda = \frac{1}{2m}$ is indicated on the top axis to facilitate comparison with experimental length scales. The red exclusion curve represents the upper bound $N_{\rm max}$ derived from Yukawa envelope constraints using Eq.~\eqref{eq:Nmax-massive}, with the shaded region above this curve excluded by current data. Horizontal dashed lines mark the massless limits from Eqs.~\eqref{eq:Ns-num}--\eqref{eq:Nv-num}, providing direct comparison between massive and massless scenarios.

Several key features emerge from this analysis. First, the massless limit provides the most stringent constraints across all particle types, with vector bosons offering the tightest bound at $N_v \lesssim 9 \times 10^{60}$ degrees of freedom. This enhanced sensitivity reflects the long-range nature of massless corrections, which scale as $1/r^3$ and remain unsuppressed at the distance scales probed by precision experiments.

For massive particles, the constraints exhibit strong scale dependence. Torsion balance experiments operating at millimeter to submillimeter scales ($r_\star \sim 10^{-4}$--$10^{-2}$ m) provide the most restrictive bounds, limiting multiplicities to $N \lesssim 10^{64}$ for particles with masses in the $10^{-16}$--$10^{-13}$ eV range. This dominance arises because torsion balances probe precisely the distance scales where massive corrections transition from the short-range exponentially suppressed regime to the intermediate regime where quantum effects are maximally observable.

At larger distances (smaller masses), the constraints weaken considerably as only planetary dynamics and lunar laser ranging (LLR) provide sensitivity. These techniques, while remarkable in their precision over astronomical scales, cannot match the controlled laboratory environment of torsion balance experiments for constraining quantum gravitational effects. The vertical bands in Figure~\ref{fig:N_massive} clearly delineate these complementary experimental regimes, illustrating how different techniques carve out distinct regions of parameter space.

The steep rise in $N_{\rm max}$ toward smaller masses reflects the exponential suppression of Yukawa corrections: as particle masses decrease, larger multiplicities are required to generate observable deviations from Newton's law. However, this trend is ultimately cut off by the massless limit, where the analytic $1/r^3$ corrections provide the fundamental theoretical ceiling.

These results demonstrate that precision gravitational tests already impose non-trivial constraints on dark sector model building. While the numerical bounds may appear large---reaching $\mathcal{O}(10^{64})$ in some cases---they represent genuine, model-independent limits on the complexity of hidden sectors that can exist while remaining consistent with laboratory measurements. As experimental sensitivity continues to improve, particularly in the submillimeter regime where torsion balance techniques excel, these bounds will tighten proportionally, providing increasingly powerful probes of beyond-Standard-Model physics through purely gravitational interactions.

The complementarity between different experimental approaches highlights an underappreciated synergy in fundamental physics: laboratory-scale precision gravity experiments, originally conceived to test general relativity or search for extra dimensions, emerge as sensitive probes of dark sector multiplicity and structure. This connection opens new avenues for constraining theoretical models that predict large numbers of light degrees of freedom, from strongly coupled dark gauge theories to string theory compactifications with extensive moduli sectors.

\section{Discussion and Conclusions}\label{sec:discussionsconclusions}

In this work we have systematically derived and analyzed quantum gravitational corrections to Newton's law arising from large dark sectors, translating effective field theory predictions into experimentally testable constraints. Our analysis demonstrates that while individual particles contribute negligibly to gravitational modifications, sectors containing many light degrees of freedom can generate cumulative effects that approach the sensitivity thresholds of current precision gravity experiments. For truly massless hidden sector particles, current inverse-square law tests constrain the multiplicities to remarkably large values: $N_s \lesssim 1.08 \times 10^{62}$ real scalars, $N_f \lesssim 1.80 \times 10^{61}$ Dirac fermions, and $N_v \lesssim 9.02 \times 10^{60}$ vectors. These bounds, while numerically enormous, represent the first direct laboratory constraints on the size of massless dark sectors coupled purely gravitationally. The $1/r^3$ power-law corrections from massless fields provide the strongest leverage for such constraints, as they are not exponentially suppressed at the distance scales probed by torsion balance experiments.

For massive dark particles, the quantum corrections take the form of Yukawa-suppressed modifications with characteristic range $\lambda = 1/(2m)$. By mapping these corrections onto the experimental Yukawa template, we have shown how published fifth-force exclusion curves can be directly translated into multiplicity bounds $N_{\rm max}(\lambda)$ for particles of mass $m = 1/(2\lambda)$. This approach provides a systematic framework for constraining the mass spectrum and abundance of hidden sector states using existing gravitational data. Beyond the specific numerical bounds, our work establishes a general methodology for connecting quantum field theory predictions in gravity with precision experimental constraints. The translation between effective field theory nonlocal form factors and the phenomenological parametrizations used by experimentalists provides a crucial bridge between theoretical dark sector model-building and laboratory tests.

The gravitational constraints derived here occupy a unique position in the landscape of dark sector searches. Unlike collider experiments, which require non-gravitational interactions, or cosmological observations, which are sensitive to the collective energy density, precision gravity tests can probe purely gravitationally-coupled sectors at the level of individual particle multiplicities. Cosmological constraints on relativistic degrees of freedom, typically expressed through effective neutrino number $N_{\rm eff}$, provide complementary information but probe different physics. While cosmology constrains the total energy contribution of light species during nucleosynthesis and recombination, precision gravity tests are sensitive to the detailed spectrum and multiplicity structure of dark sectors at much later times and smaller scales. The mass ranges accessible to laboratory gravity experiments, corresponding to Compton wavelengths from micrometers to millimeters, complement both high-energy collider searches and astrophysical probes. This intermediate scale regime is particularly relevant for dark sectors arising from string theory compactifications or strongly-coupled hidden gauge theories, where light states with masses in the meV to eV range naturally emerge.

Note that our bounds are laboratory and model-independent: they constrain the \emph{present-day} multiplicity of light states without assuming a cosmological population. Cosmology can provide stronger constraints when the hidden sector was populated and remained relativistic at BBN/CMB epochs, typically phrased as limits on $\Delta N_{\rm eff}$. However, low reheating temperatures, suppressed portals, late entropy injection, or purely gravitational/out-of-equilibrium production histories can substantially weaken cosmological sensitivity while leaving our reinterpretation intact. We therefore view the two approaches as complementary: CMB/BBN constrain \emph{abundances} under specified histories, whereas precision gravity constrains the \emph{spectrum} of light degrees of freedom irrespective of their relic density.

Our results have direct implications for several classes of theoretical models. Strongly coupled dark gauge sectors may contain multiple light states (e.g.\ PNGBs or glueballs) below the experimental lever arm; while each contributes negligibly, the cumulative effect scales with the number of such light states: any unexpectedly rich light spectrum is directly testable in the same, model-independent way. Our bounds provide quantitative guidance for the maximum complexity such sectors can achieve while remaining consistent with laboratory tests. Extra-dimensional theories often predict towers of Kaluza-Klein modes or large numbers of moduli fields, and the gravitational constraints complement existing bounds from fifth-force searches to help restrict the allowed parameter space of phenomenologically viable compactifications. Similar analyses could be extended to axion-like particles that couple to gravity through topological terms or induce effective modifications to gravitational couplings.

 It is important to remark that Kaluza–Klein towers fit naturally into our framework {\it provided one sums only over modes lighter than the probed scale}. In the short-distance regime, massive KK gravitons generate a set of Yukawa deviations whose envelope is already captured by our mapping; the well-studied short-range gravity limits in ADD/RS scenarios can be rephrased in exactly this language. Our plots should thus be read as providing an efficient, model-agnostic translation between measured inverse-square-law tests and the cumulative effect of any tower of light modes.

We also note that for extremely large $N$, loop-induced higher-derivative poles can drift down toward the laboratory window. In that regime, our reinterpretation remains applicable and typically becomes \emph{more} constraining, because such poles manifest as Yukawa deviations from $1/r$ that are bounded by the same precision data.

Several aspects of our analysis merit further investigation. We have focused on leading-order quantum corrections arising from one-loop effects, but higher-order contributions could modify the detailed form of the gravitational modifications, particularly in strongly-coupled dark sectors where perturbative calculations may break down. Our treatment assumes minimal gravitational coupling through the stress-energy tensor, but dark sector particles with non-minimal couplings to curvature could generate different signatures that might be more readily accessible to experiment. The analysis presented here applies to vacuum quantum corrections, though in cosmological or astrophysical contexts, thermal corrections from dark sector particles could modify the gravitational dynamics in observable ways. The constraints presented here are based on current experimental sensitivities, and ongoing improvements in torsion balance experiments, lunar laser ranging, and other precision gravity tests will tighten these bounds and potentially access new regions of parameter space.

The intersection of precision gravity experiments with dark sector phenomenology represents an underexplored frontier in fundamental physics. While the Standard Model contributions to gravitational quantum corrections are utterly negligible, the situation changes qualitatively in the presence of large hidden sectors. This work demonstrates that laboratory tests of Newton's law, originally conceived to search for extra dimensions or light scalar fields, can serve as powerful probes of the multiplicity and structure of dark matter sectors. The enormous numerical values of the bounds we derive, reaching $\mathcal{O}(10^{61})$ for some particle types, should not obscure their physical significance. These constraints represent genuine, model-independent limits on the complexity of dark sectors that could exist in nature while remaining consistent with precision tests of gravity. As experimental sensitivity improves, these bounds will tighten proportionally, providing increasingly stringent tests of theoretical models that predict large numbers of light degrees of freedom.

We have established a quantitative framework for constraining large dark sectors using precision tests of Newton's gravitational law. By systematically translating quantum field theory predictions for loop-induced gravitational corrections into the experimental language of Yukawa deviations and inverse-square law violations, we derive robust bounds on the multiplicities and mass spectra of hidden sector particles coupled only gravitationally. Our results demonstrate that current torsion balance experiments and related precision gravity tests already impose non-trivial constraints on dark sector model building, with bounds reaching $\mathcal{O}(10^{61})$ for massless degrees of freedom. As experimental techniques continue to improve, this approach will provide increasingly powerful tests of theories predicting large numbers of light particles beyond the Standard Model. The methodology developed here opens new avenues for connecting theoretical dark sector phenomenology with laboratory experiments, highlighting an unexpected synergy between precision gravity measurements and particle physics beyond the Standard Model, and establishing precision gravitational tests as a valuable complement to collider searches and cosmological observations in the quest to understand the hidden sectors that may populate our universe.

\begin{acknowledgments}
This work is partly supported by the U.S.\ Department of Energy grant number de-sc0010107 (SP). 
\end{acknowledgments}

\bibliographystyle{unsrtnat}
\bibliography{references}

\end{document}